


\font\twelverm=cmr10  scaled 1200   \font\twelvei=cmmi10  scaled 1200
\font\twelvesy=cmsy10 scaled 1200   \font\twelveex=cmex10 scaled 1200
\font\twelvebf=cmbx10 scaled 1200   \font\twelvesl=cmsl10 scaled 1200
\font\twelvett=cmtt10 scaled 1200   \font\twelveit=cmti10 scaled 1200
\font\twelvesc=cmcsc10 scaled 1200
\skewchar\twelvei='177   \skewchar\twelvesy='60


\def\twelvepoint{\normalbaselineskip=12.4pt plus 0.1pt minus 0.1pt
  \abovedisplayskip 12.4pt plus 3pt minus 9pt
  \belowdisplayskip 12.4pt plus 3pt minus 9pt
  \abovedisplayshortskip 0pt plus 3pt
  \belowdisplayshortskip 7.2pt plus 3pt minus 4pt
  \smallskipamount=3.6pt plus1.2pt minus1.2pt
  \medskipamount=7.2pt plus2.4pt minus2.4pt
  \bigskipamount=14.4pt plus4.8pt minus4.8pt
  \def\rm{\fam0\twelverm}          \def\it{\fam\itfam\twelveit}%
  \def\sl{\fam\slfam\twelvesl}     \def\bf{\fam\bffam\twelvebf}%
  \def\mit{\fam 1}                 \def\cal{\fam 2}%
  \def\sc{\twelvesc}		   \def\tt{\twelvett}
  \def\sf{\twelvesf}
  \textfont0=\twelverm   \scriptfont0=\tenrm   \scriptscriptfont0=\sevenrm
  \textfont1=\twelvei    \scriptfont1=\teni    \scriptscriptfont1=\seveni
  \textfont2=\twelvesy   \scriptfont2=\tensy   \scriptscriptfont2=\sevensy
  \textfont3=\twelveex   \scriptfont3=\twelveex  \scriptscriptfont3=\twelveex
  \textfont\itfam=\twelveit
  \textfont\slfam=\twelvesl
  \textfont\bffam=\twelvebf \scriptfont\bffam=\tenbf
  \scriptscriptfont\bffam=\sevenbf
  \normalbaselines\rm}



\def\beginlinemode{\endmode
  \begingroup\parskip=0pt \obeylines\def\\{\par}\def\endmode{\par\endgroup}}
\def\beginparmode{\endmode
  \begingroup \def\endmode{\par\endgroup}}
\let\endmode=\par
{\obeylines\gdef\
{}}
\def\singlespace{\baselineskip=\normalbaselineskip}

\def\oneandahalfspace{\baselineskip=\normalbaselineskip
  \multiply\baselineskip by 3 \divide\baselineskip by 2}
\def\doublespace{\baselineskip=\normalbaselineskip \multiply\baselineskip by 2}

\newcount\firstpageno
\firstpageno=2
\footline={\ifnum\pageno<\firstpageno{\hfil}\else{\hfil\twelverm\folio\hfil}\fi}
\def\toppageno{\global\footline={\hfil}\global\headline
  ={\ifnum\pageno<\firstpageno{\hfil}\else{\hfil\twelverm\folio\hfil}\fi}}
\let\rawfootnote=\footnote		
\def\footnote#1#2{{\rm\singlespace\parindent=0pt\parskip=0pt
  \rawfootnote{#1}{#2\hfill\vrule height 0pt depth 6pt width 0pt}}}
\def\raggedcenter{\leftskip=4em plus 12em \rightskip=\leftskip
  \parindent=0pt \parfillskip=0pt \spaceskip=.3333em \xspaceskip=.5em
  \pretolerance=9999 \tolerance=9999
  \hyphenpenalty=9999 \exhyphenpenalty=9999 }
\def\dateline{\rightline{\ifcase\month\or
  January\or February\or March\or April\or May\or June\or
  July\or August\or September\or October\or November\or December\fi
  \space\number\year}}
\def\received{\vskip 3pt plus 0.2fill
 \centerline{\sl (Received\space\ifcase\month\or
  January\or February\or March\or April\or May\or June\or
  July\or August\or September\or October\or November\or December\fi
  \qquad, \number\year)}}


\hsize=6.5truein
\hoffset=0.0truein
\vsize=8.9truein
\voffset=0.0truein
\parskip=\medskipamount
\toppageno
\twelvepoint
\doublespace
\def\\{\cr}
\overfullrule=0pt 




\def\title#1{			
   \null \vskip 3pt plus 0.3fill \beginlinemode
   \doublespace \raggedcenter {\bf #1} \vskip 3pt plus 0.1 fill}

\def\author			
  {\vskip 3pt plus 0.1fill \beginlinemode \doublespace \raggedcenter}

\def\affil			
  {\vskip 3pt \beginlinemode \doublespace \raggedcenter \it}

\def\abstract			
  {\vskip 3pt plus 0.1fill \subhead {Abstract:}
   \beginparmode \narrower \oneandahalfspace }

\def\endtopmatter		
  {\vskip 3pt plus 0.1fill \endpage \body}

\def\body			
  {\beginparmode}		

\def\head#1{			
   \goodbreak \vskip 0.4truein	
  {\immediate\write16{#1} \raggedcenter {\bf #1} \par}
   \nobreak \vskip 3pt \nobreak}

\def\subhead#1{			
  \vskip 0.25truein		
  {\raggedcenter {\it #1} \par} \nobreak \vskip 3pt \nobreak}

\def\beneathrel#1\under#2{\mathrel{\mathop{#2}\limits_{#1}}}

\def\refto#1{${\,}^{#1}$}	

\newdimen\refskip \refskip=0pt
\def\references		
  {\head{References}	
   \beginparmode \frenchspacing \parindent=0pt \leftskip=\refskip
   \parskip=0pt \everypar{\hangindent=20pt\hangafter=1}}

\gdef\refis#1{\item{#1.\ }}			

\gdef\journal#1, #2, #3 {		
    {\it #1}, {\bf #2}, #3.}		

\def\refstylePR{ 			
  \gdef\refto##1{${\,}^{##1}$}		
  \gdef\refis##1{${\,}^{##1}$}		
  \gdef\journal##1, ##2, ##3, ##4 {	
    \rm ##1 {\bf ##2}, ##3 (##4).}}



\def\prd{\journal Phys. Rev. D}
\def\prl{\journal Phys. Rev. Lett.}

\def\np{\journal Nucl. Phys.}


\def\endreferences{\body}

\def\figurecaptions		
  {\endpage \beginparmode \head{Figure Captions}
   \parskip=3pt \everypar{\hangindent=20pt\hangafter=1} }

\def\endpage			
  {\vfill\eject}

\def\endpaper	{\endmode\vfill\supereject}
\def\endjnl	{\endpaper\end}


\def\ref#1{ref.{#1}}			
\def\Ref#1{Ref.{#1}}			
\def\[#1]{[\cite{#1}]}
\def\cite#1{{#1}}


\def\(#1){(\call{#1})}
\def\call#1{{#1}}
\def\frac#1#2{{#1 \over #2}}

\def\12{{1\over2}}

\def\sla{\raise.15ex\hbox{$/$}\kern-.57em}
\def\leaderfill{\leaders\hbox to 1em{\hss.\hss}\hfill}
\def\twiddle{\lower.9ex\rlap{$\kern-.1em\scriptstyle\sim$}}
\def\bigtwiddle{\lower1.ex\rlap{$\sim$}}
\def\gtwid{\mathrel{\raise.3ex\hbox{$>$\kern-.75em\lower1ex\hbox{$\sim$}}}}
\def\ltwid{\mathrel{\raise.3ex\hbox{$<$\kern-.75em\lower1ex\hbox{$\sim$}}}}
\def\square{\kern1pt\vbox{\hrule height 1.2pt\hbox{\vrule width 1.2pt\hskip 3pt
   \vbox{\vskip 6pt}\hskip 3pt\vrule width 0.6pt}\hrule height 0.6pt}\kern1pt}
\def\tdot#1{\mathord{\mathop{#1}\limits^{\kern2pt\ldots}}}

\def\pmb#1{\setbox0=\hbox{#1}%
  \kern-.025em\copy0\kern-\wd0
  \kern  .05em\copy0\kern-\wd0
  \kern-.025em\raise.0433em\box0 }

\def\del {\nabla}



\refstylePR

\catcode`@=11
\newcount\r@fcount \r@fcount=0
\newcount\r@fcurr
\immediate\newwrite\reffile
\newif\ifr@ffile\r@ffilefalse
\def\w@rnwrite#1{\ifr@ffile\immediate\write\reffile{#1}\fi\message{#1}}

\def\writer@f#1>>{}
\def\referencefile{
  \r@ffiletrue\immediate\openout\reffile=\jobname.ref%
  \def\writer@f##1>>{\ifr@ffile\immediate\write\reffile%
    {\noexpand\refis{##1} = \csname r@fnum##1\endcsname = %
     \expandafter\expandafter\expandafter\strip@t\expandafter%
     \meaning\csname r@ftext\csname r@fnum##1\endcsname\endcsname}\fi}%
  \def\strip@t##1>>{}}

\def\citeall#1{\xdef#1##1{#1{\noexpand\cite{##1}}}}
\def\cite#1{\each@rg\citer@nge{#1}}	

\def\each@rg#1#2{{\let\thecsname=#1\expandafter\first@rg#2,\end,}}
\def\first@rg#1,{\thecsname{#1}\apply@rg}	
\def\apply@rg#1,{\ifx\end#1\let\next=\relax
\else,\thecsname{#1}\let\next=\apply@rg\fi\next}

\def\citer@nge#1{\citedor@nge#1-\end-}	
\def\citer@ngeat#1\end-{#1}
\def\citedor@nge#1-#2-{\ifx\end#2\r@featspace#1 
  \else\citel@@p{#1}{#2}\citer@ngeat\fi}	
\def\citel@@p#1#2{\ifnum#1>#2{\errmessage{Reference range #1-#2\space is bad.}
    \errhelp{If you cite a series of references by the notation M-N, then M and
    N must be integers, and N must be greater than or equal to M.}}\else%
 {\count0=#1\count1=#2\advance\count1
by1\relax\expandafter\r@fcite\the\count0,%
  \loop\advance\count0 by1\relax
    \ifnum\count0<\count1,\expandafter\r@fcite\the\count0,%
  \repeat}\fi}

\def\r@featspace#1#2 {\r@fcite#1#2,}	
\def\r@fcite#1,{\ifuncit@d{#1}		
    \expandafter\gdef\csname r@ftext\number\r@fcount\endcsname%
    {\message{Reference #1 to be supplied.}\writer@f#1>>#1 to be supplied.\par
     }\fi%
  \csname r@fnum#1\endcsname}

\def\ifuncit@d#1{\expandafter\ifx\csname r@fnum#1\endcsname\relax%
\global\advance\r@fcount by1%
\expandafter\xdef\csname r@fnum#1\endcsname{\number\r@fcount}}

\let\r@fis=\refis			
\def\refis#1#2#3\par{\ifuncit@d{#1}
    \w@rnwrite{Reference #1=\number\r@fcount\space is not cited up to now.}\fi%
  \expandafter\gdef\csname r@ftext\csname r@fnum#1\endcsname\endcsname%
  {\writer@f#1>>#2#3\par}}

\def\r@ferr{\endreferences\errmessage{I was expecting to see
\noexpand\endreferences before now;  I have inserted it here.}}
\let\r@ferences=\references
\def\references{\r@ferences\def\endmode{\r@ferr\par\endgroup}}

\let\endr@ferences=\endreferences
\def\endreferences{\r@fcurr=0
  {\loop\ifnum\r@fcurr<\r@fcount
    \advance\r@fcurr by 1\relax\expandafter\r@fis\expandafter{\number\r@fcurr}%
    \csname r@ftext\number\r@fcurr\endcsname%
  \repeat}\gdef\r@ferr{}\endr@ferences}


\let\r@fend=\endpaper\gdef\endpaper{\ifr@ffile
\immediate\write16{Cross References written on []\jobname.REF.}\fi\r@fend}

\catcode`@=12

\citeall\refto		
\citeall\ref		%
\citeall\Ref		%

\title {Black string traveling waves}

\author {David Garfinkle}

\affil {Dept. of Physics, Oakland University, Rochester, MI 48309}

\abstract
A family of solutions to low energy string theory is found.  These
solutions represent waves traveling along ``extremal black strings.''

\endtopmatter

\head {1. Introduction}

The low energy limit of string theory gives a set of classical equations,
similar to Einstein's equations, for a metric and other fundamental fields.
Several solutions of these equations have been found, including some
representing ``black strings''\refto{HS} that is extended objects with
horizons.  In this paper we will find some new solutions to low energy string
theory.  The solutions are found using a generating technique; starting with
a known solution the technique produces a new solution representing waves
traveling on the old background spacetime.  There is no restriction on the
amplitude of the waves and, in particular, no approximation in which the waves
are considered ``small'' is used.  The background solutions we will use are
black strings; so the new solutions represent waves traveling on a black string
background.

The fields in the low energy string theory lagrangian are the sigma model
metric
$ {{\bar g}_{ab}} , $
the dilaton
$ \phi $
and the axion field strength
$ {H_{abc}} . \; $
Here
$ {H_{abc}} $
is derived from a potential
$ B_{ab} $
by
$ {\bf H} = d {\bf B} . \; $
The action in
$ D $
dimensions is
$$
S = \int \; {d^D} x \; {\sqrt {\bar g}} \; {e^\phi } \; \left [ {\bar R}
\; + \; {\del ^a} \phi {\del _a} \phi \; - \; {1 \over {12}} \;
{H_{abc}} {H^{abc}} \right ] \; \; \; .
\eqno(1)
$$
Here
$ \bar R $
is the scalar curvature of
$ {{\bar g}_{ab}} $
and all indicies are raised and lowered with
$ {{\bar g}_{ab}} . \; $

To use the generating technique we will need to write our expressions in terms
of the Einstein metric
$ g_{ab} $
rather than the sigma model metric
$ {{\bar g}_{ab}} . \; $
Define the number
$ n , $
scalar
$ \psi $
and metric
$ g_{ab} $
by
$$
n \equiv D \; - \; 4 \; \; \; ,
\eqno(2)
$$
$$
\psi \equiv {\phi \over {n \, + \, 2}} \; \; \; ,
\eqno(3)
$$
$$
{g_{ab}} \equiv {e^{2 \psi }} \; {{\bar g}_{ab}} \; \; \; .
\eqno(4)
$$
In terms of these variables the action is
$$
S = \int {d^{n + 4}} x \; {\sqrt g} \; \left [ R \; - \; (n + 2 ) {\del
^a} \psi {\del _a} \psi \; - \; {1 \over {12}} \; {e^{4 \psi }} \;
{H_{abc}} {H^{abc}} \right ] \; \; \; .
\eqno(5)
$$
Here, and in all subsequent equations, indicies are raised and lowered with the
Einstein metric
$ {g_{ab}} . \; $
{}From this action it follows that the equations for the fields are
$$
{\del ^a} {\del _a} \psi \; - \; {1 \over {6 ( n + 2 )}} \; {e^{4 \psi
}} \; {H_{abc}} {H^{abc}} = 0 \; \; \; ,
\eqno(6)
$$
$$
{\del _a} \left ( {e^{4 \psi }} \; {H^{abc}} \right ) = 0 \; \; \; ,
\eqno(7)
$$
$$
{{R^a}_b} = (n + 2) {\del ^a} \psi {\del _b} \psi \; + \; {1 \over 4} \;
{H^{acd}} {H_{bcd}}
$$
$$
- \; {1 \over {6 (n + 2)}} \; {{\delta ^a}_b} \; {H^{cde}} {H_{cde}} \;
\; \; .
\eqno(8)
$$

We will find solutions of equations (6-8) using the traveling wave generating
technique of reference\cite{GV} .  The technique works as follows: let
$ ({g_{ab}} , \psi , {B_{ab}} ) $
be a solution of equations (6-8) that has a null, hypersurface orthogonal
Killing vector
$ {k^a} . \; $
Then there is a scalar
$ A $
such that
$$
{\del _a} {k_b} = {k_{[ a}} {\del _{b] }} \ln A \; \; \; .
\eqno(9)
$$
Define the new metric
$ {{\tilde g}_{ab}} $
by
$$
{{\tilde g}_{ab}} \equiv {g_{ab}} \; + \; A \; \Phi \; {k_a} \; {k_b} \;
\; \; .
\eqno(10)
$$
Let the scalar
$ \Phi $
satisfy the equations
$$
{k^a} {\del _a} \Phi = 0 \; \; \; ,
\eqno(11)
$$
$$
{\del ^a} {\del _a} \Phi = 0 \; \; \; .
\eqno(12)
$$
Our new solution is
$ ( {{\tilde g}_{ab}} , \psi , {B_{ab}} )$,
that is the metric
$ {\tilde g}_{ab} $
instead of
$ g_{ab} $
and the same matter fields as in the background solution.
It follows from equation (11) that
$ k^a $
is also a null Killing vector for the new solution.  So any disturbances in the
new solution must propagate at the speed of light without changing their
amplitude or shape.  That is, the new solution is a traveling wave.

One can show that
$( {{\tilde g}_{ab}} , \psi , {B_{ab}} )$
satisfies equations (6-8) as follows: a calculation of the Ricci tensor
$  {{{\tilde R}^a}_{ \; \; b}} $
of the metric
$ {{\tilde g}_{ab}} $
shows\refto{GV} that
$  {{{\tilde R}^a}_{ \; \; b}} \; - \; {{R^a}_b}  $
is proportional to
$ {\del _a} {\del ^a} \Phi . \; $
It then follows from equation (12) that
$  {{{\tilde R}^a}_{ \; \; b}} = {{R^a}_b} \, . \; $
One then needs to check that raising the indicies of
$ {H_{abc}} $
with the new metric is the same as in the old metric.  It then follows that
equation (8) is satisfied.  Now the wave operator
$ {\del _a} {\del ^a} $
for scalars invariant under the null translation
is the same in the new metric as in the old; so equation (6) is satisfied.
Similarly the divergence of an antisymmetric tensor is the same in both
metrics; so equation (7) is satisfied.

Therefore all one needs to do, in order to produce a new solution, is solve
equations (11) and (12) on the old spacetime.  In order to apply this technique
the old spacetime must have a null Killing vector.
The ``extremal black string'' solutions of reference \cite{HS}
have a null Killing vector.  In the
next section we will apply the generating technique to these solutions.

\head {2. Traveling waves in $D \ge 5 $ dimensions }

The extremal black string solutions\refto{HS, HHS} in
$ D \ge 5 $
dimensions are given by the following expressions:
$$
\psi = {1 \over {n + 2}} \; \ln \left ( 1 \; + \; {M \over {r^n}} \right
) \; \; \; ,
\eqno(13)
$$
$$
{\bf B} = \left ( 1 \; - \; {e^{ - ( n + 2 ) \psi }} \right ) d u \wedge
d v \; \; \; ,
\eqno(14)
$$
$$
d {s^2} = 2 {e^{ - n \psi }} d u d v \; + \; {e^{2 \psi }} \; \left (
d {r^2} \; + \; {r^2} d {\Omega ^2 _{n + 1}} \right ) \; \; \; .
\eqno(15)
$$
Here
$ d {\Omega ^2 _{n + 1 }} $
is the metric of the unit
$ n + 1 $
sphere and
$ M $
is a constant.

The null Killing vector is
$$
{k^a} = {{\left ( {\partial \over {\partial v}} \right ) }^a}
\eqno(16)
$$
and the scalar
$ A $
is
$ A = {e^{n \psi }} . $

We first verify that
$ H_{abc} $
with the indicies raised by the new metric is the same as with the old metric.
The inverse of the new metric is
$$
{{\left ( {{\tilde g}^{ - 1}} \right ) }^{ab}} = {g^{ab}} \; - \; A \Phi
{k^a} {k^b} \; \; \; .
\eqno(17)
$$
{}From equations (14) and (16) it follows that there is an
$ s_a $
such that
$ {k^a} {s_a} = 0 $
and
$$
{k^a} {H_{abc}} = {s_{[b}} {k_{c]}} \; \; \; .
\eqno(18)
$$
It then follows from equation (17) that
$ H_{abc} $
with its indicies raised by the new metric is the same as
$ {H^{abc}} . $

The new metric is given by
$$
d {{\tilde s}^2} = d {s^2} \; + \; {e^{ - n \psi }} \; \Phi \; d {u^2}
\eqno(19)
$$
where
$ \Phi $
satisfies equations (11) and (12).  Equation (11) is just
$$
{{\partial \Phi } \over {\partial v}} = 0 \; \; \; .
\eqno(20)
$$
so
$ \Phi $
depends only on the coordinates
$ u , \; r $
and the
$ n + 1  $
sphere coordinates.  We seek a solution of the form
$$
\Phi = f ( u ) \; P ( r ) \; Y ( {\theta _i} )
\eqno(21)
$$
where the
$ \theta _i $
are
the coordinates on the
$ n + 1 $
sphere.  Equation (12) then becomes
$$
0 = {e^{ - 2 \psi }} \; f \; \left [ {Y \over {r^{n + 1}}} \; {d \over
{d r}} \; \left ( {r^{n + 1}} \; {{d P } \over {d r}} \right ) \; + \;
{P \over {r^2}} \; {D^2} Y \right ] \; \; \; .
\eqno(22)
$$
Here
$ D^2 $
is the Laplacian operator on the
$ n + 1 $
sphere.  This equation places no restrictions on
$ f ( u ) . \; $
Essentially the function
$ f $
is the profile of the wave and can be chosen arbitrarily.  From equation (22)
it follows that
$ Y $
must be an eigenfunction of
$ {D^2} . \; $
The eigenvalues of
$ D^2 $
on the
$ n + 1 $
sphere are
$ - \, l ( l + n ) $
where
$ l $
is a nonnegative integer.  The function
$ Y $
is then
$ {Y_l} , $
an
$ n + 1 $
spherical harmonic corresponding to the integer
$ l . \; $
It then follows that
$ P $
satisfies the equation
$$
{1 \over {r^{n + 1 }}} \; {d \over d r} \; \left ( {r^{n + 1}} \; {{d P}
\over {d r}} \right ) \; - \; {{l ( l + n )} \over {r^2}} \; P = 0 \; \;
\; .
\eqno(23)
$$
The solutions are
$ P = {r^l} $
or
$ P = {r^{ - ( l + n )}} . \; $
The metric is then
$$
d {{\tilde s}^2} = {e^{ - n \psi }} \; \left ( 2 \, d u \, d v \; + \;
f ( u ) \; {r^\beta } \; {Y_l} ( {\theta _i} ) \; d {u^2} \right )
$$
$$
+ \; {e^{ 2 \psi }} \; \left ( d {r^2} \; + \; {r^2} \, d {\Omega ^2 _{n
+ 1}} \right )
\eqno(24)
$$
where
$ \beta = l $
or
$ \beta = - \, ( l + n ) . $

\head {3. Traveling waves in $ 3 $ dimensions }

The extremal black string in
$ 3 $
dimensions is given by\refto{HHS}
$$
\psi = \ln \left ( {\sqrt k} \, r \right )
\eqno(25)
$$
where
$ k $
is a constant;
$$
{\bf B} = {M \over r} \; d u \wedge d v \; \; \; ,
\eqno(26)
$$
$$
d {s^2} = 2 \, k \, {r^2} \, \left ( 1 \; - \; {M \over r} \right ) \; d
u \, d v \; + \; {1 \over 4} \; {k^2} \; {{\left ( 1 \; - \; {M \over r}
\right ) }^{ - 2}} \; d {r^2} \; \; \; .
\eqno(27)
$$
Once again the null Killing vector is
$ {k^a} = {{( \partial / \partial v )}^a} . \; $
The scalar
$ A $
is now
$ A = {{\left [ {k^2} ( {r^2} \, - \, M r ) \right ] }^{ - 1 }} \; . \; $
As before, the three form
$ H_{abc} $
is the same with its indicies raised by the new metric as with the old metric.
The new metric is
$$
d {{\tilde s}^2} = d {s^2} \; + \; {r^2} \; \left ( 1 \; - \; {M \over
r} \right ) \; \Phi \; d {u^2} \; \; \; .
\eqno(28)
$$
Equation (11) tells us that
$ \Phi $
depends only on
$ r $
and
$ u . \; $
We seek a solution of the form
$$
\Phi = f ( u ) \; P ( r ) \; \; \; .
\eqno(29)
$$
Equation (12) then becomes
$$
0 = f ( u ) \; {d \over {d r}} \; \left [ {{( r \, - \, M )}^2} \;
{{d P} \over {d r}} \right ] \; \; \; .
\eqno(30)
$$
As before, this equation places no restrictions on
$ f ( u ) . \; $
The solutions for
$ P ( r ) $
are
$ P = 1 $
or
$ P = {{( r - M )}^{ - 1}} \; . \; $
The first solution gives a new metric equivalent to the old metric.  So the
only physically distinct solution is
$$
d {{\tilde s}^2} = k \, {r^2} \, \left ( 1 \; - \; {M \over r} \right )
\; d u \; d v \; + \; r \; f ( u ) \; d {u^2} \; + \; {1 \over 4} \;
{k^2} \; {{\left ( 1 \; - \; {M \over r} \right ) }^{ - 2 }} \; d {r^2}
\; \; \; .
\eqno(31)
$$

\head {4. Properties of the metrics}

The extremal black strings are singular at
$ r = 0 . \; $
Therefore the traveling wave solutions are also singular at
$ r = 0 . \; $
One might interpret the extremal black string as the field of a straight
fundamental string.  In that case the traveling wave solutions would represent
propagating modes of the fundamental string.

Some three dimensional black string solutions come from exact conformal field
theories.\refto{HH}  It would be interesting to see whether there is
an exact conformal
field theory whose low energy limit is the three dimensional black
string traveling wave.

\head {Acknowledgement}

I would like to thank Gary Horowitz and Tanmay Vachaspati for
helpful discussions.

\references

\refis{HS} G. Horowitz and A. Strominger, \np , B360, 197, 1991

\refis{GV} D. Garfinkle and T. Vachaspati, \prd , 42, 1960, 1990

\refis{HHS} J. Horne, G. Horowitz and A. Steif, \prl , 68, 568, 1992

\refis{HH} J. Horne and G. Horowitz, \np , B368, 444, 1992

\endreferences

\endjnl
\end